\documentclass[12pt,preprint]{aastex}

\newcommand {\hI} {\ion{H}{1}\,\,}
\newcommand {\hII} {\ion{H}{2}\,\,}
\newcommand {\nII} {\ion{N}{2}\,\,}
\newcommand {\ha} {H$\alpha$\,\,}
\newcommand {\kms} {\,km\,s$^{-1}$\,}
\newcommand {\M} {\mbox{${\cal M}$}}
\newcommand {\E} {\mbox{${\cal E}$}}
\newcommand {\msol} {\M$_\odot$\,}
\newcommand {\lsol} {L$_\odot$\,}
\newcommand {\mlb} {(\M/L$_B$)$_\star$\,\,}
\newcommand {\mlsol}{\mbox{${\cal M}_\odot$/L$_{\odot}$}}
\newcommand{\PA}{$\mathrm{PA}$}
\newcommand{\FM} {\texttt{\textsc{FaNTOmM}}}
\newcommand{\um} {$\mu$m\,}

\slugcomment{Accepted for Publication in {\it The Astronomical Journal} }

\shorttitle{NGC 7793 - Deep \ha observations}
\shortauthors{Dicaire et al.}

\begin{document}

\title{Deep Fabry-Perot \ha Observations of NGC 7793:\\
a Very Extended \ha Disk and a Truly Declining Rotation Curve}

\author{I. Dicaire\altaffilmark{1}, C. Carignan\altaffilmark{1},
P. Amram\altaffilmark{2}, M. Marcelin\altaffilmark{2},
J. Hlavacek--Larrondo\altaffilmark{1},\\
M.--M. de Denus-Baillargeon\altaffilmark{1},
O. Daigle\altaffilmark{1,2} and O. Hernandez\altaffilmark{1}}

\altaffiltext{1}{Laboratoire d'Astrophysique Exp\'erimentale,
D\'epartement de physique, Universit\'e de Montr\'eal, C. P. 6128,
Succ. centre-ville, Montr\'eal, Qu\'e., Canada H3C 3J7}
\altaffiltext{2}{Laboratoire d'Astrophysique de Marseille,
Observatoire Astronomique Marseille Provence, Universit\'e de
Provence \& CNRS, 2 place Le Verrier, F-13248 Marseille CEDEX 04,
France}

\begin{abstract}
Deep \ha observations of the Sculptor Group galaxy NGC 7793
were obtained on the ESO 3.60m and the Marseille
36cm telescopes at La Silla, Chile. \ha emission is detected
all the way to the edge of the \hI disk, making of the \hII
disk of NGC 7793 one of the largest ever observed in a quiet
non-AGN late--type system. Even in the very outer parts, the
\hII ionizing sources are probably mainly internal (massive
stars in the disk) with an unlikely contribution from the
extragalactic ionizing background. The \ha kinematics confirms
what had already been seen with the \hI observations: NGC 7793
has a truly declining rotation curve. However, the decline is not Keplerian
and a dark halo is still needed to explain the rotation velocities
in the outer parts.
\end{abstract}


\section{Introduction}

NGC 7793 is one of the 5 bright members of the classical Sculptor Group
along with NGC 55, NGC 247, NGC 253 and NGC 300. Sculptor is the closest
group of galaxies outside the Local Group. It covers an area of $\sim 20\degr$
in diameter centered on the constellation Sculptor at $\alpha = 0^h30^m$
and $\delta = -30\degr$ \citep{dev59}.
However, it is still debated whether Sculptor is really a group. \citet{kar03}
suggest instead that it may be part of a filament that extends along the
line of sight from the Local Group out to $\sim 5$ Mpc.

The optical parameters of NGC 7793 are summarized in Table~\ref{opt_par}.
It is a typical Sd galaxy with a very tiny bulge and a filamentary
spiral structure (see figs 1 \& 2).
It is an intrinsically small system
with an exponential scale length $\alpha^{-1} \simeq 1$ kpc. For this study,
a distance of 3.38 Mpc \citep{puc88} is adopted, based on
numerous distance indicators. However, somewhat larger distances of e.g.
3.82 \citep{kar03} and 3.91 Mpc \citep{kar05} have also been suggested.
For our adopted distance, NGC 7793 has an absolute B magnitude of --18.3
for a total blue luminosity of $\sim 3.1 \times 10^9$ \lsol.

\begin{table}
\caption{Optical parameters of NGC 7793.\label{opt_par}}
\begin{tabular}{llc}
\tableline\tableline
Parameter & Value & Reference \\
\tableline
Morphological type & SA(s)d & a \\
R.A. (2000) & 23{\rm $^h$} 57{\rm$^m$} {49\fs5} & a \\
Dec. (2000) & --32\arcdeg\ 35\arcmin\ 24\arcsec & a \\
Isophotal major diameter $D_{25}$ & 10.1$\arcmin$ & b \\
Holmberg radius $R_{HO}$ & 6.1$\arcmin$ & b \\
Exponential scale length $\alpha^{-1}$, kpc & 1.1 & b \\
Axis ratio $q = b/a$ & 0.60 & b \\
Inclination ($q_0 = 0.12$), $i$ & 53.7$\degr$ & b \\
Position angle, \PA & 279.3$\degr$  & b \\
Corrected total $B$ magnitude, $B{^{0,i}_T}$ ($A_g$ = 0.02) & 9.33 & a,b \\
Adopted distance (Mpc) & 3.38 & c \\
& (1$\arcmin$ = 0.98 kpc) & \\
Absolute $B$ magnitude, $M^{0,i}_B$ & --18.31 & b \\
Total blue luminosity ($M_{\odot}$ = 5.43), \lsol &
$3.1 \times 10^9$ & b \\
(\bv) & 0.54 & a \\
\tableline\tableline
\end{tabular}
\tablenotetext{a}{\citet{dev91}}
\tablenotetext{b}{\citet{car85a}}
\tablenotetext{c}{\citet{puc88}}
\end{table}

Previous kinematical studies of NGC 7793 were done in the optical,
using Fabry-Perot interferometry by \citet{dav80}. Their \ha
rotation curve (RC) extends to $\sim 4 \arcmin$, which barely
reaches the maximum velocity. Ten years later, deep VLA \hI
observations (13.5 hours in C/D configuration) allowed one to derive
the rotation curve twice as far, out to $\sim 8 \arcmin$
\citep{car90}. A remarkable result of those \hI observations was
that, contrary to most spirals, the rotation curve is not flat in
the outer parts but appears to be declining ($\Delta V_{rot} \simeq
30$ \kms or 25\% of $V_{max}$ between the maximum velocity and the
last point of the RC), even after a careful modeling (tilted--ring
model) of the warp ($\Delta$\PA $\simeq 20 \degr$) in the outer \hI
disk. While the RC is declining in the outer parts, it is less steep
than a pure Keplerian decline and a dark halo is still needed to
properly model the mass distribution.

It would be interesting to be able to check the uniqueness of this
RC with independent observations using another tracer than \hI. As
we have seen, previous \ha observations \citep{dav80} only extend to
$\sim 4 \arcmin$ or 80\% of the optical radius $R_{25}$. However,
deep \ha imaging with a 68 \AA\  (FWHM) filter \citep{fer96} showed
that the diffuse ionized gas (DIG, also referred to as warm ionized
medium, WIM) component in NGC 7793 easily extended out to $R_{25}$.
Thus, one of the main motivations of the observations presented in
this paper was the hope to extend the \ha kinematics, at least out
to that point, in order to confirm (or not) the decline of the \hI
RC.

The other motivation was to try to detect that DIG component out to
very low levels, if possible at radii larger than $R_{25}$. Large
DIG components have been seen in nearby starburst and active
galaxies \citep{vei03} and can sometimes have sizes comparable to
the \hI component (see e.g. NGC 1068), but not much is known on the
extent of the DIG component of quiet late--type systems.
Comprehensive studies of the large scale structure of DIG components
have been done on edge--on systems \citep{ran90, det92, vei95}, on
selected small regions of M31 \citep{wal94}, and on large samples
\citep{ken95, zur00, thi02, meu06, oey07}.

Massive stars clearly provide the largest source of Lyman continuum
(Lyc) photons in spiral galaxies; an issue concerns whether the bulk
of these photons are deposited over localized regions, such as the
Str\"omgren spheres which define \hII regions, or whether a
significant fraction can escape from the regions of recent star
formation where they were created and thus ionize the interstellar
medium over much larger scales \citep{fer96}. The large scale radial
distribution of the DIG across galactic disks can provide stringent
constraints on the source of its ionization. For example, if the
number of Lyc photons produced in star--forming regions were the
only factor responsible for producing the DIG, then the ionized gas
distribution should follow very closely, both on small and large
scales, the distribution of discrete \hII regions. This addresses
the possible internal sources of ionization.

However, if ionized gas is detected past the \hI truncation edge
\citep{cor89, van93}, one possible explanation for that truncation
could be that, at large radii, the thin \hI disks get fully ionized
by the metagalactic UV background, as discussed
by \citet{bla97}. This is another incentive to try to push the
detection of the diffuse \ha as far out as possible. If \ha could be
detected that far, it could even be used to put a limit on that UV background.

Section II will present the new Fabry-Perot \ha observations
and discuss data reduction while Section III will derive the kinematical
parameters and the rotation curves.
This will be followed in Section IV by a study of the
mass distribution and Section V will discuss the extent of the
DIG. Finally, the principal results will be discussed in Section VI and
a summary and the main conclusions will be given in Section VII.

\section{Observations \& Data Reduction}

The Fabry-Perot (FP) observations were obtained on the Marseille
36cm (from 2005 October 27 to 2005 November 2) and on the ESO 3.6m
(2005 November 3) telescopes at La Silla, Chile, using the photon
counting camera \FM\ \citep{gac02, her03}. The same focal reducer,
interference filter, Fabry-Perot etalon and camera were used on both
telescopes. The interference filter, with a peak transmission
$T_{max} \sim$ 80\%, was centered at $\lambda_c = 6563$\AA\ and had
a FWHM of 30.4\AA\,. The interference order of the Fabry-Perot
interferometer was p = 765 at \ha for a free spectral range (FSR) of
8.6\AA\ or $\sim$390 \kms. The mean \textit{Finesse} of the etalon
was $\sim 18.5$ (resolution $\simeq 14 000$). The finesse is a
dimensionless value expressing the spectral resolution of the etalon
($\Delta\lambda = \frac{FSR (\AA)}{Finesse}$). The FSR was scanned
in 60 channels by steps of 0.14\AA. The photon counting camera \FM\
is based on a Hamamatsu photocathode coupled with a Dalsa commercial
CCD. The photocathode has a quantum efficiency of $\sim$23\% at \ha
and the CCD has 1024x1024 12.5\um\, square pixels. The CCD was
operated in its low spatial resolution where pixels are binned 2x2
at 40 frames per second.

A photon counting camera, such as \FM, is an essential tool for this
kind of work. Its ability to rapidly scan the FP interferometer
allows the photometric variations to be averaged out. For
comparison, in CCD observations, each FP channel must be observed
for at least 5 continuous minutes to avoid the read-out noise of the
CCD from masking the weak galaxy's signal. This means that
photometric conditions must not significantly change for $\sim$5
hours with CCD observations. In photon counting, channels are
observed for 5 to 15 seconds and a complete cycle is obtained every
5 to 15 minutes. Many cycles are made during an observation and
since the data are analyzed on--line,  SNR estimations can be made
throughout the observations and the observer can decide when to stop
the integration. The wavelength calibration is done using a Ne lamp
at 6598.95\AA. Since the calibration lamp is strong compared to
galaxies' fluxes, calibrations are done in analogic mode (non photon
counting). Typically, calibration channels are integrated 1 second
each, such that a whole calibration only takes 1 minute.

The galaxy was observed for a total of 100 minutes on the 3.60m (1.67
minutes/channel) and for 1200 minutes (20 minutes/channel) over 4
nights on the 36cm telescope.
Raw observational data consist of many data files that contain photons'
positions for every cycle/channel duo. With a cycling of 10 seconds
integration time per channel, one file was created every 10 seconds.
The different steps of the data reduction are:
\begin{itemize}
\item integration of raw data files into an interferogram data cube
(3D data cube sliced for every Fabry-Perot channel);
\item phase correction of interferograms to create wavelength-sorted data
cubes (3D data cube sliced for every wavelength interval);
\item hanning spectral smoothing;
\item sky emission removal;
\item spatial binning/smoothing (e.g. adaptive binning);
\item radial velocity map extraction from emission line positions;
\item addition of astrometry information;
\item kinematical information extraction.
\end{itemize}

All the reduction was performed with IDL routines inspired of the
ADHOCw reduction package
(http://www.oamp.fr/adhoc/adhoc/adhocw.htm). Details on the data
reduction can be found in \citet{her05}, \citet{che06} and
\citet{dai06a} and the routines are available at
http://www.astro.umontreal.ca/$\sim$odaigle/reduction. Especially,
the way the profiles are found, the way the sky emission is
subtracted, which is a crucial step when trying to detect very weak
emission, the way the data is smoothed using an adaptive smoothing
technique are all described in details in \citet{dai06b}.

\begin{figure}
\begin{center}
  \includegraphics[height=0.8\textheight]{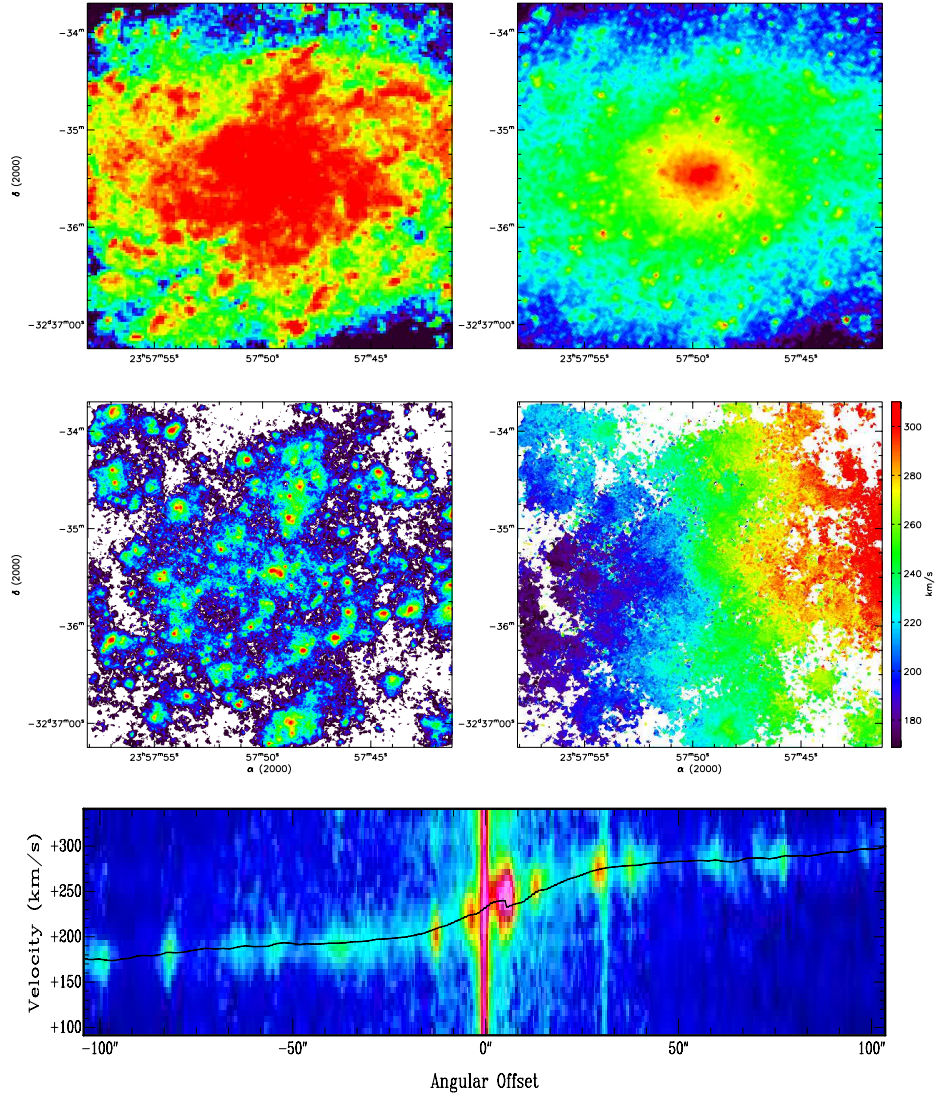}
\caption{Fabry-Perot observations of NGC 7793 on the 3.60m.
\textbf{Top left} : DSS Blue Band image.
\textbf{Top right} : SPITZER IRAC 3.6\um image.
\textbf{Middle left} : \ha monochromatic image.
\textbf{Middle right} : \ha velocity field.
\textbf{Bottom} : PV diagram.}
\label{obs3m60}
\end{center}
\end{figure}
\clearpage

\begin{figure}
\begin{center}
  \includegraphics[height=0.8\textheight]{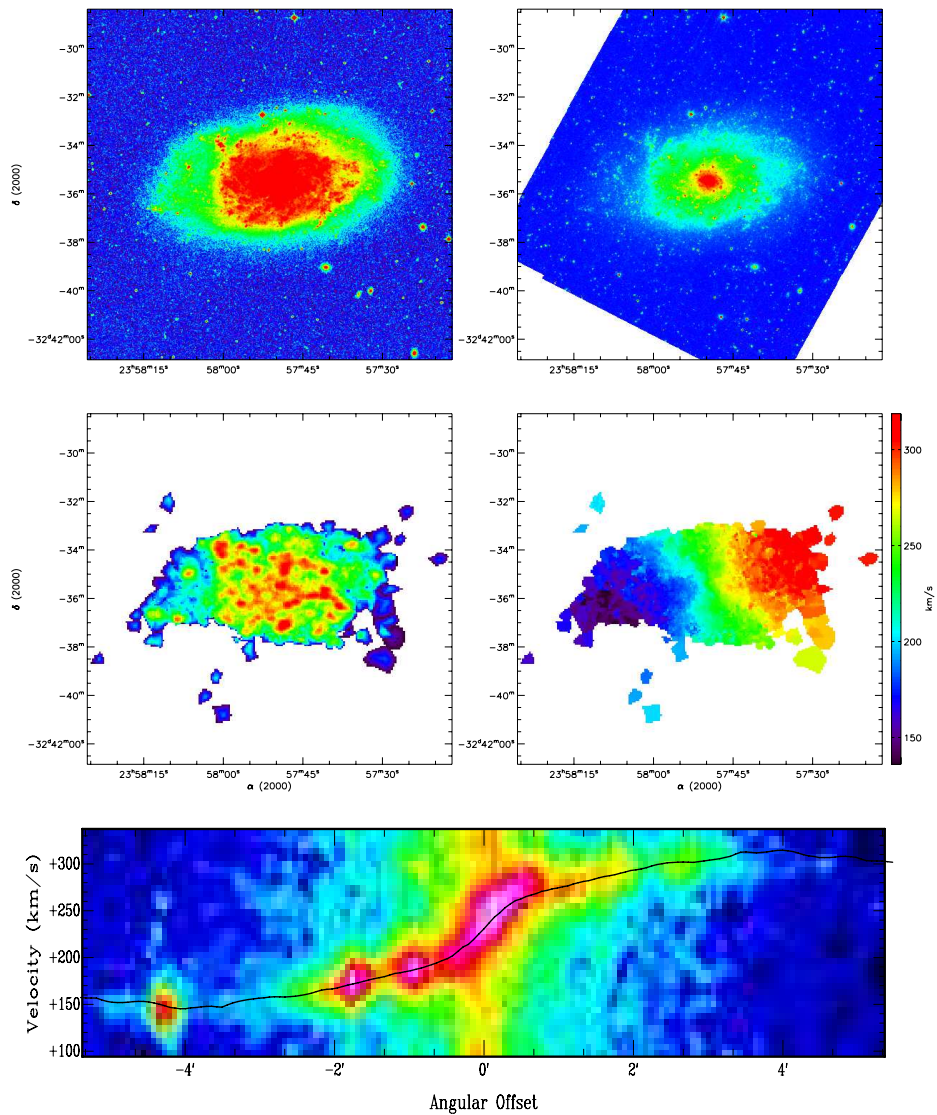}
\caption{Fabry-Perot observations of NGC 7793 on the 36cm.
\textbf{Top left} : DSS Blue Band image.
\textbf{Top right} : SPITZER IRAC 3.6\um image.
\textbf{Middle left} : \ha monochromatic image.
\textbf{Middle right} : \ha velocity field.
\textbf{Bottom} : PV diagram.}
\label{obs36cm}
\end{center}
\end{figure}
\clearpage

A first attempt at detecting the DIG in the outer disk of NGC 7793
was done in October 1993, during 13 observing nights, for a
cumulative total exposure time of 64 hours. The same instrumentation
was used (36cm telescope, focal reducer, interference filter and
interferometer), but the detector was different. This first data set
was obtained with our old IPCS, a Thompson camera based on a
Silicium Nocticon TV tube technology \citep{bou84}. However, the
camera used to obtain the data set presented in this paper used a
new GaAs tube technology \citep{gac02} and has a detective quantum
efficiency $\sim$25\%, at least four times larger than the old
Thompson IPCS. Thus, this second run was less time consuming (20
hours instead of 64) for reaching a higher SNR. Moreover, the
shortening of the individual elementary exposure time together with
the diminution of the total observing time and better reduction
software \citep{dai06b} allowed a better removal of the night sky
lines which is critical at such a low signal level.

\section{Kinematical Parameters and Rotation Curves}

In order to derive the rotation curves, one must find the set of
orientation parameters (rotation center ($x_0, y_0$), systemic
velocity $V_{sys}$, inclination $i$, position angle of the major axis \PA)
that best represents the observed velocity fields, shown in
fig.~\ref{obs3m60} and fig.~\ref{obs36cm}, at all radii.
The data in an opening angle of 40\degr\ about the minor axis
are excluded from the analysis to minimize errors due to deprojection
effects. A cosine weighting function (using the angle $\theta$ from
the major axis) is used for the rest of the data, which gives
maximum weighting on the major axis. The task \textit{ROTCUR}
in the reduction package \textit{GIPSY} \citep{vog01} is used to find
those parameters.

Since they are correlated, the dynamical center and the systemic
velocity are looked for first, by keeping $i$ and \PA\ fixed (using
the photometric values given in Table~\ref{opt_par}). We find that
($x_0, y_0$) corresponds to the optical center and that $V_{sys}$ =
238 $\pm 3$ \kms for the 3.60m and 230 $\pm 4$ \kms for the 36cm
data. This can be compared to the \citet{car90} \hI global profile
midpoint velocity of 230 $\pm 2$ \kms and intensity-weighted mean
velocity of 235 $\pm 4$ \kms or to their systemic velocity derived
in the same way from the \hI velocity field of 227 $\pm 3$ \kms. The
agreement is very good considering that a slight zero point shift
can be expected since the calibration is done at the Ne wavelength
and not at the observed wavelength.

\begin{figure}
\plottwo{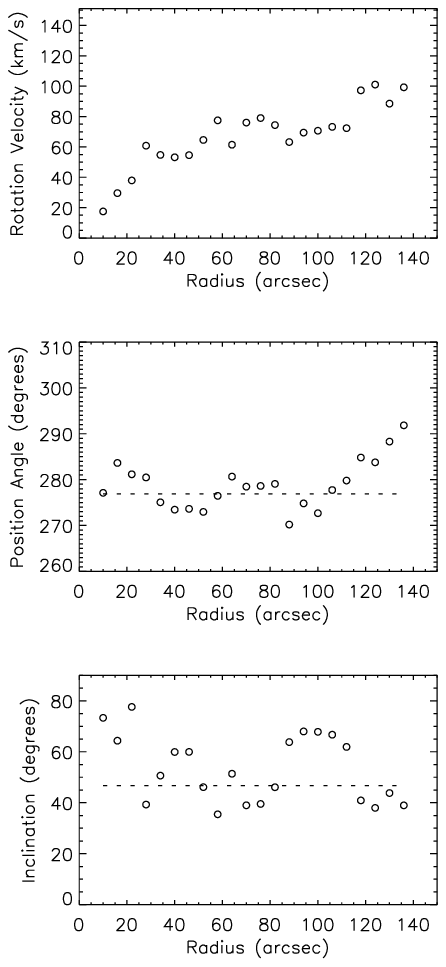}{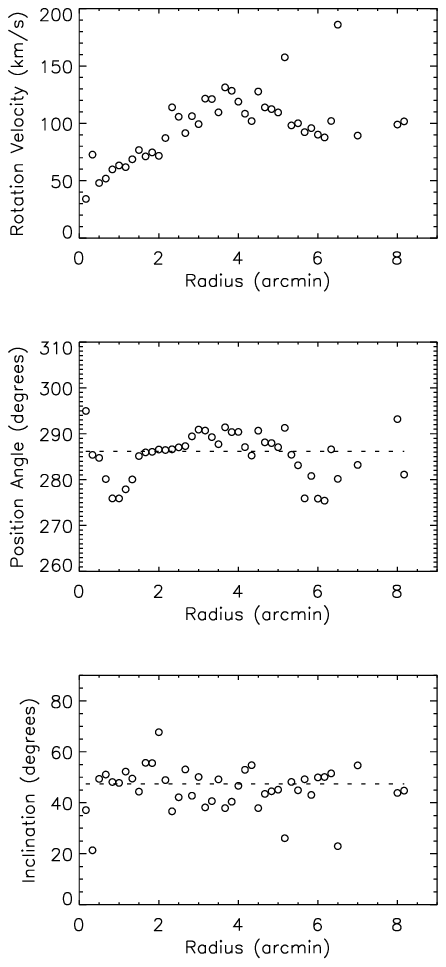}
\caption{Kinematical parameters of NGC 7793 derived from the
3.60m (\textbf{left}) and the 36cm (\textbf{right}) Fabry-Perot observations.
\label{KP}}
\end{figure}

\begin{figure}
\plottwo{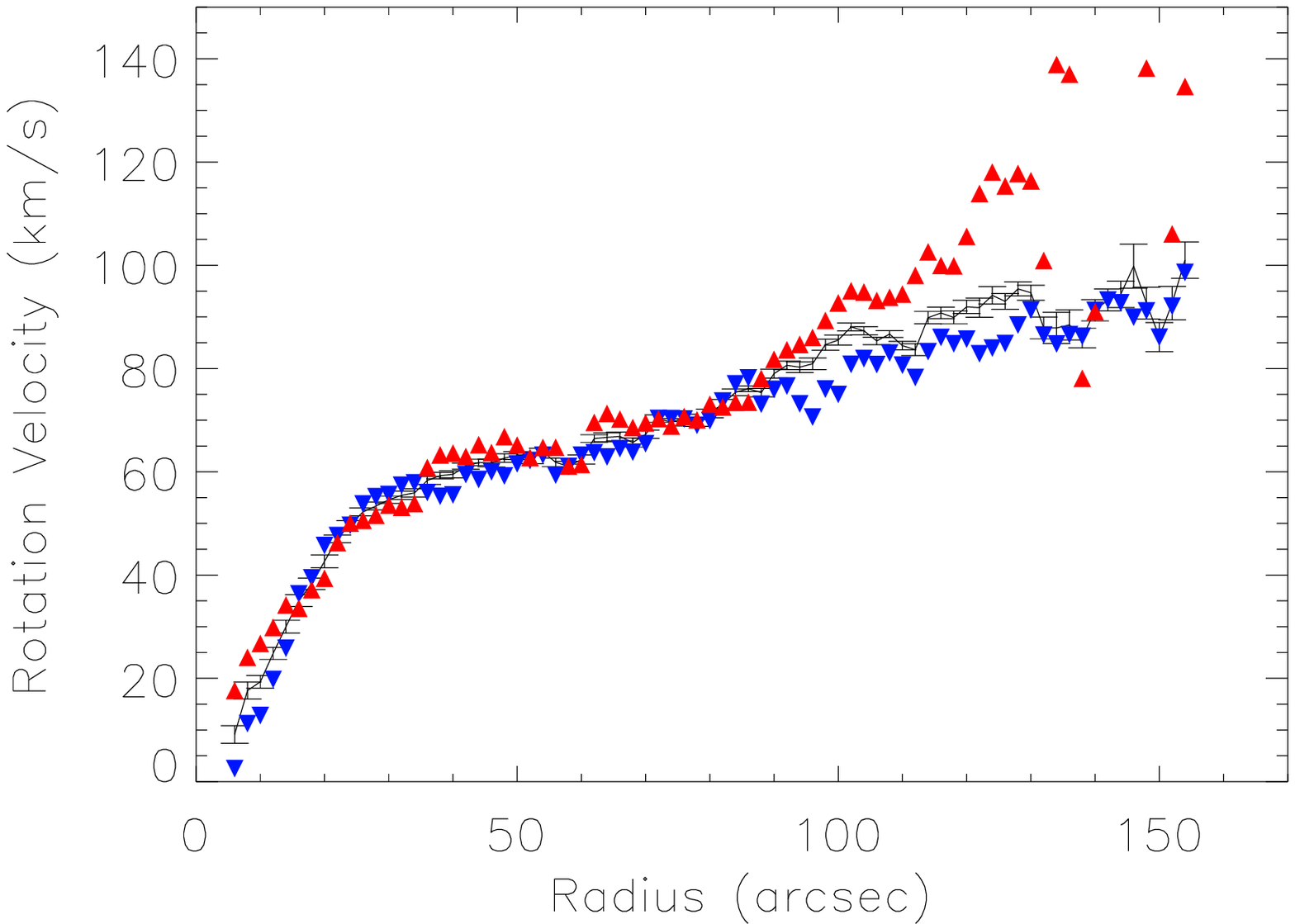}{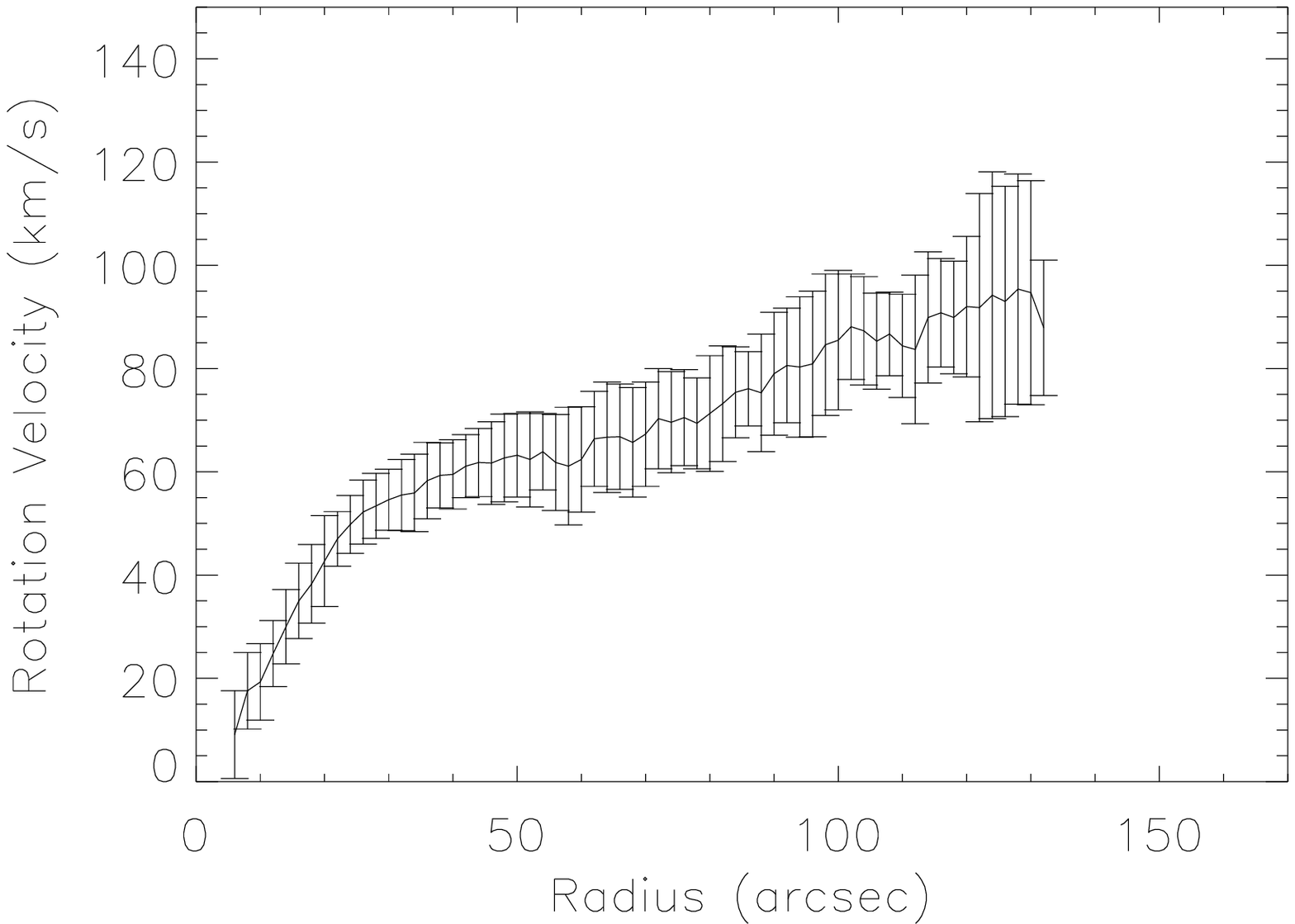} \caption{\textbf{Left} :
Rotation curve derived from the 3.60m data for the whole galaxy
(black) and separately for the receding (red) and the approaching
(blue) sides with V$_{sys}$ = 238 \kms, \PA = 277\degr\ and i =
47\degr. \textbf{Right} : 3.60m adopted \ha rotation curve for NGC
7793. The data are also tabulated in Table~\ref{RC3m60}.
\label{vrot3m60}}
\end{figure}

\begin{figure}
\plottwo{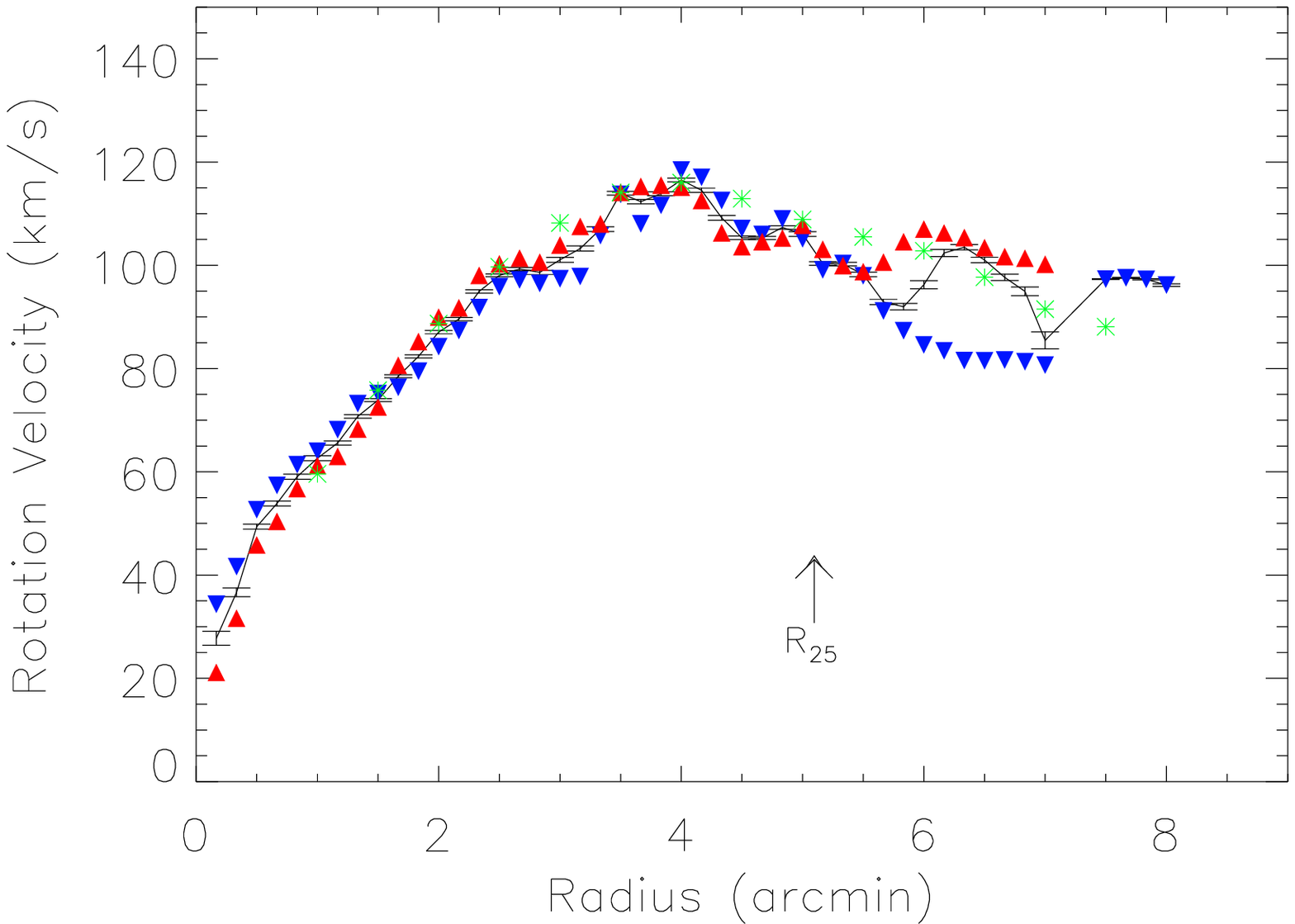}{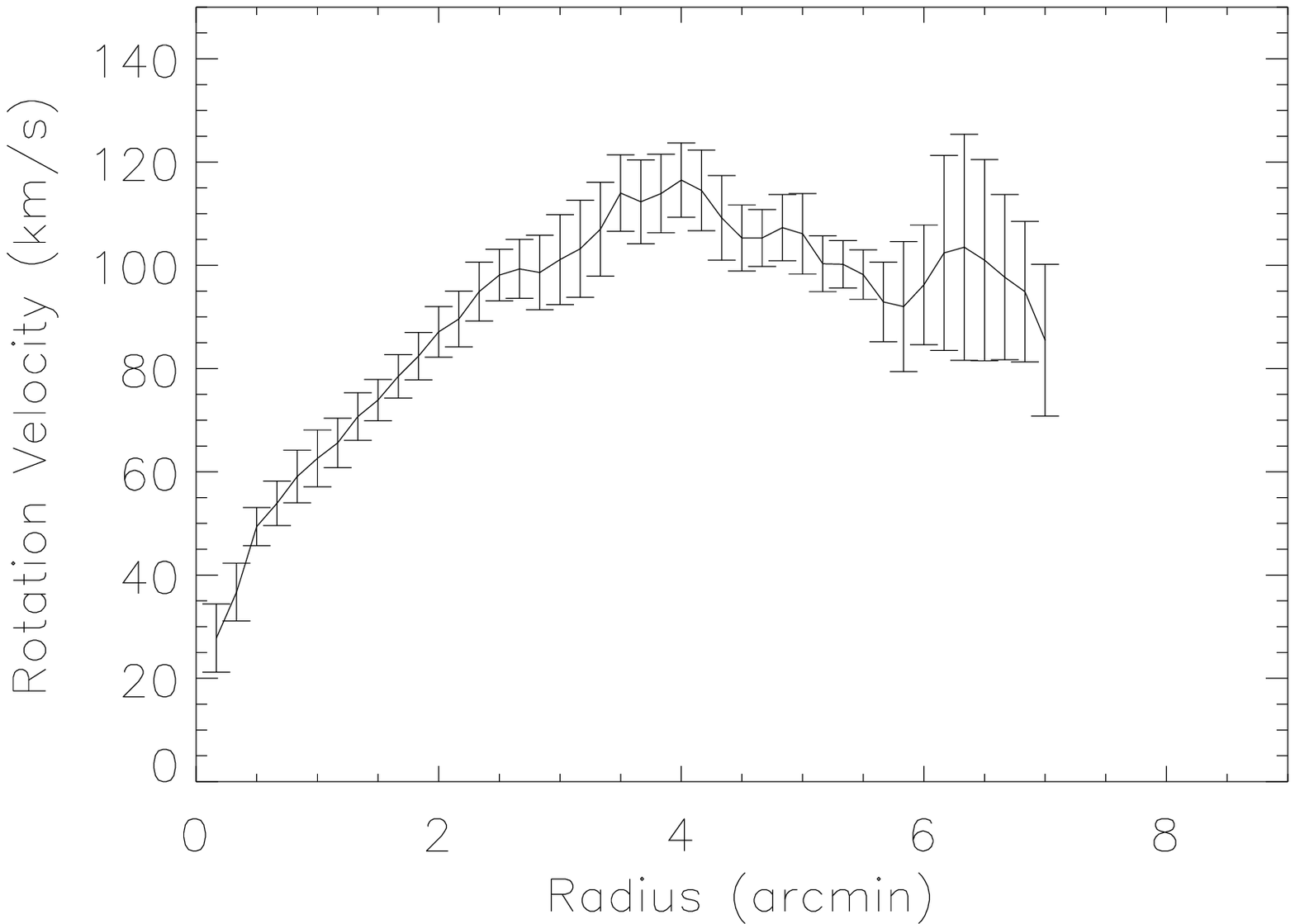} \caption{\textbf{Left} :
Rotation curve derived from the 36cm data for the whole galaxy
(black) and separately for the receding (red) and the approaching
(blue) sides with V$_{sys}$ = 230 \kms, \PA = 286\degr\ and i =
47\degr. The \hI data are plotted as crosses (green). \textbf{Right}
: 36cm adopted \ha rotation curve for NGC 7793. The data are also
tabulated in Table~\ref{RC36cm}. \label{vrot36cm}}
\end{figure}

The next step is to obtain a least--squares solution for $i$, \PA\
and $V_{rot}$ in concentric annuli in the plane of the galaxy, by
keeping ($x_0, y_0$) and $V_{sys}$ fixed. This is shown in Fig.~\ref{KP}.
For both data sets,  an inclination $i$ = 47\degr\
$\pm 9$\degr\ for the 3.6m and $\pm 6$\degr\ for the 36cm is adopted.
For the \PA\ we found 277\degr\ $\pm 3$\degr\ for the 3.60m
and 286\degr\ $\pm 4$\degr\ for the 36 cm.
This is consistent with the \hI data \citep{car90} which
show that the \PA\ is smaller in the inner parts than in the outer parts.
In order to see if the solution found is a good representation of
the whole galaxy, separate solutions are also obtained for the
approaching and the receding sides. The 3 solutions (both sides,
approaching, receding) and the adopted rotation curves are given
in Fig.~\ref{vrot3m60} for the 3.60m and in Fig.~\ref{vrot36cm} for the
36cm data. The
adopted errors are the biggest difference between the kinematical
solution for both sides and the separate solutions for either the approaching
or the receding sides or the intrinsic error from the tilted-ring
model, if larger. The values of the RCs are tabulated
in Table~\ref{RC3m60} and Table~\ref{RC36cm}, respectively.

For the 3.60m data of Fig.~\ref{vrot3m60}, only the velocity points up to
132\arcsec\ will be used because further out there are not enough
points on the receding side and the errors become too large.
Similarly, for the 36cm data in Fig.~\ref{vrot36cm},
even if the \ha emission is detected out to
$\sim$8.5\arcmin\ ($\sim R_{HI}$), only the data up to 7 \arcmin\ will be
used because there is no emission after that radius on the
receding side.

\begin{table}
\caption{\ha RC from the 3.60m data.\label{RC3m60}}
\begin{tabular}{cccccc}
\hline\hline
Radius & $V_{\rm rot}$ & $\Delta V_{\rm rot}$ &
Radius & $V_{\rm rot}$ & $\Delta V_{\rm rot}$ \\
(arcsec) & (\kms) & (\kms) & (arcsec) & (\kms) & (\kms) \\
\hline
6.0 & 9.1 & 8.5 & 70.0 & 67.3  & 10.1 \\
8.0 & 17.6 & 7.4 & 72.0 & 70.3 & 9.7 \\
10.0 & 19.3 & 7.4 & 74.0 & 69.6 & 9.8 \\
12.0 & 24.8 & 6.4 & 76.0 & 70.5 & 9.3 \\
14.0 & 30.0 & 7.2 & 78.0 & 69.4 & 8.8 \\
16.0 & 35.0 & 7.3 & 80.0 & 71.3 & 11.2 \\
18.0 & 38.3 & 7.6 & 82.0 & 73.2 & 11.2 \\
20.0 & 42.7 & 8.8 & 84.0 & 75.4 & 8.8 \\
22.0 & 47.0 & 5.3 & 86.0 & 76.1 & 7.2 \\
24.0 & 49.8 & 5.6 & 88.0 & 75.3 & 11.4 \\
26.0 & 52.2 & 6.2 & 90.0 & 79.0 & 11.9 \\
28.0 & 53.4 & 6.3 & 92.0 & 80.6 & 11.1 \\
30.0 & 54.6 & 5.9 & 94.0 & 80.3 & 13.6 \\
32.0 & 55.5 & 6.9 & 96.0 & 80.9 & 14.1 \\
34.0 & 55.9 & 7.5 & 98.0 & 84.6 & 13.7 \\
36.0 & 58.3 & 7.4 & 100.0 & 85.5 & 13.5 \\
38.0 & 59.3 & 6.3 & 102.0 & 88.1 & 10.2 \\
40.0 & 59.5 & 6.7 & 104.0 & 87.3 & 10.5 \\
42.0 & 61.1 & 6.1 & 106.0 & 85.3 & 9.3 \\
44.0 & 61.8 & 6.6 & 108.0 & 86.7 & 8.1 \\
46.0 & 61.7 & 8.0 & 110.0 & 84.4 & 10.0 \\
48.0 & 62.7 & 8.5 & 112.0 & 83.7 & 14.4 \\
50.0 & 63.2 & 8.1 & 114.0 & 89.9 & 12.7 \\
52.0 & 62.4 & 9.2 & 116.0 & 90.8 & 10.5 \\
54.0 & 63.9 & 7.4 & 118.0 & 89.9 & 10.9 \\
56.0 & 61.8 & 9.3 & 120.0 & 92.0 & 13.6 \\
58.0 & 61.1 & 11.4 & 122.0 & 91.8 & 22.1 \\
60.0 & 62.4 & 10.2 & 124.0 & 94.2 & 23.9 \\
62.0 & 66.4 & 9.2 & 126.0 & 93.0 & 22.3 \\
64.0 & 66.7 & 10.7 & 128.0 & 95.4 & 22.3 \\
66.0 & 66.8 & 10.2 & 130.0 & 94.7 & 21.7 \\
68.0 & 65.7 & 10.6 & 132.0 & 87.9 & 13.1 \\
\hline
\end{tabular}
\end{table}

\begin{table}
\caption{\ha RC from the 36cm data.\label{RC36cm}}
\begin{tabular}{cccccc}
\hline\hline
Radius & $V_{\rm rot}$ & $\Delta V_{\rm rot}$ &
Radius & $V_{\rm rot}$ & $\Delta V_{\rm rot}$ \\
(arcsec) & (\kms) & (\kms) & (arcsec) & (\kms) & (\kms) \\
\hline
10.0 & 27.8 & 6.6 & 220.0 & 112.3 & 8.1 \\
20.0 & 36.7 & 5.6 & 230.0 & 113.9 & 7.6 \\
30.0 & 49.4 & 3.7 & 240.0 & 116.5 & 7.2 \\
40.0 & 53.9 & 4.3 & 250.0 & 114.5 & 7.8 \\
50.0 & 59.1 & 5.1 & 260.0 & 109.2 & 8.2 \\
60.0 & 62.6 & 5.5 & 270.0 & 105.3 & 6.4 \\
70.0 & 65.6 & 4.8 & 280.0 & 105.3 & 5.5 \\
80.0 & 70.7 & 4.6 & 290.0 & 107.3 & 6.4 \\
90.0 & 73.9 & 4.0 & 300.0 & 106.1 & 7.8 \\
100.0 & 78.5 & 4.2 & 310.0 & 100.3 & 5.4 \\
110.0 & 82.4 & 4.6 & 320.0 & 100.2 & 4.6 \\
120.0 & 87.1 & 4.9 & 330.0 & 98.2 & 4.8 \\
130.0 & 89.6 & 5.4 & 340.0 & 92.9 & 7.7 \\
140.0 & 94.9 & 5.7 & 350.0 & 92.0 & 12.6 \\
150.0 & 98.1 & 5.0 & 360.0 & 96.2 & 11.6 \\
160.0 & 99.3 & 5.7 & 370.0 & 102.4 & 18.9 \\
170.0 & 98.6 & 7.2 & 380.0 & 103.5 & 21.9 \\
180.0 & 101.1 & 8.7 & 390.0 & 101.0 & 19.5 \\
190.0 & 103.2 & 9.4 & 400.0 & 97.7 & 16.0 \\
200.0 & 107.0 & 9.1 & 410.0 & 94.9 & 13.6 \\
210.0 & 114.0 & 7.4 & 420.0 & 85.5 & 14.7 \\
\hline
\end{tabular}
\end{table}

\section{Mass Modeling}

\begin{figure}
%\centering
%\includegraphics[width=\columnwidth]{vrotfinal.eps}
\plotone{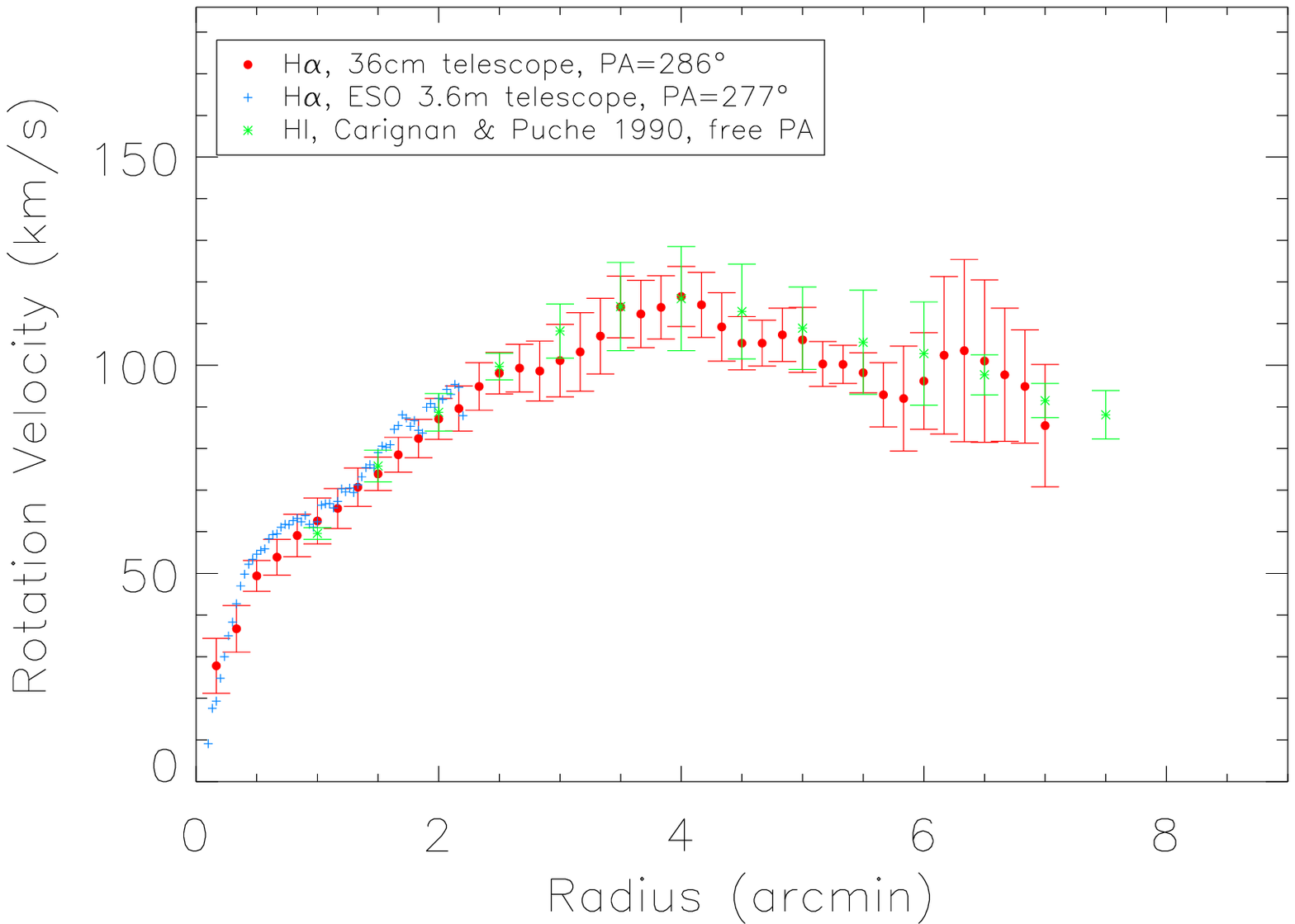} \caption{\hI (green), 3.60m \ha (blue) and
36cm \ha (red) rotations curves of NGC 7793.} \label{vrotfinal}
\end{figure}

As can be seen in Fig.~\ref{vrotfinal}, the \ha RC is very similar
to the \hI RC (errors on the 3.60m data were not plotted to
facilitate the comparison; see Fig.~\ref{vrot3m60}) so that the mass
models are expected to be very similar. However, since the \hI
velocities are slightly lower in the very inner parts and slightly
higher in the intermediate parts, one could expect slight
differences.

In order to compare our results with those obtained using the \hI RC
\citep{car90}, the same approach for the study of the mass
distribution will be used. The dark halo is represented by an
isothermal sphere which can be described by two free parameters: the
one-dimensional velocity dispersion $\sigma$ and the core radius
$r_c$. The central density is then given by $\rho_0 = 9 \sigma^2 / 4
\pi G r_c^2$. The other free parameter is the mass--to--light ratio
($M / L$)$_*$ of the stellar disk. Details on the method used can be
found in \citet{car85b} and \citet{car85c}.

\begin{figure}
%\centering
%\includegraphics[width=\textwidth]{modmass.eps}
\plotone{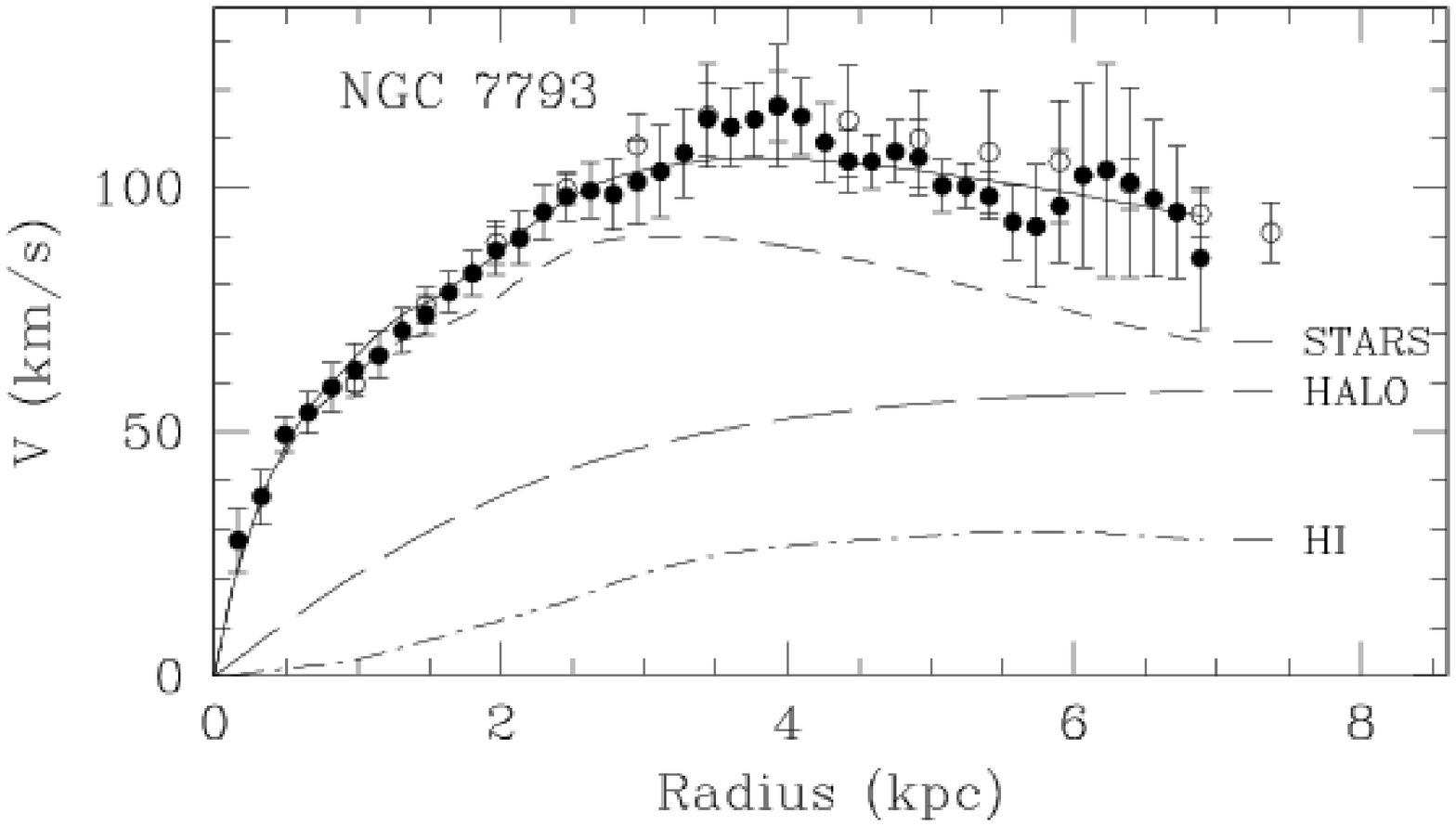} \caption{Mass model (isothermal sphere halo)
for the 36cm \ha rotation curve (filled circles) of NGC 7793. The
\hI curve (open circles) is also shown.} \label{plt_massmod}
\end{figure}

The parameters of the models and the main results are given in
Table~\ref{res_massmod} and illustrated in Fig.~\ref{plt_massmod}.
As expected, the slightly higher \ha velocity $\sim$1 kpc gives a
higher \mlb for the stellar disk, thus reducing the dark component
and the slightly lower \ha velocities $\sim$5 kpc also gives a lower
density dark component. Those two factors, combined with the fact
that the \hI RC extends slightly further out, explain the fact that
the $M_{dark} / M_{lum}$ goes down from 0.97 to 0.75 at the last
point. Nevertheless, within the uncertainties, one can say that the
\ha data confirms the results obtained with the \hI data with
$M_{dark} / M_{lum} \simeq 0.75-1.0$ at the last velocity point for
a total mass for NGC 7793 of $\sim 1.3-1.5 \times 10^{10}
M_{\odot}$. Again, one can see that even if the RC is declining over
the second half of the radius range, a dark halo is still needed to
reproduce the observed kinematics.

\begin{table}
\caption{Parameters and results for the mass models.\label{res_massmod}}
\begin{tabular}{lll}
\tableline\tableline
& \hI rotation curve$^a$ & 36cm \ha rotation curve$^b$ \\
\tableline
Luminous component & \mlb = 2.2 \mlsol & \mlb = 2.6 $\pm 0.3$ \mlsol \\
 & & \\
Dark halo component & $r_c$ = 2.7 kpc &
                      $r_c$ = 2.9 $\pm 0.8$ kpc \\
 & $\sigma$ = 40.8 \kms &
   $\sigma$ = 37.0 $\pm 6.0$\kms \\
 & $\rho_0$ = 0.038 \msol pc$^{-3}$ &
   $\rho_0$ = 0.027 $\pm 0.009$ \msol pc$^{-3}$ \\
 & & \\
At $R_{HO}$ (6.0 kpc) & $\rho_{halo}$ = 0.0025 \msol pc$^{-3}$ &
                        $\rho_{halo}$ = 0.0020 \msol pc$^{-3}$ \\
 & $M_{dark} / M_{lum}$ = 0.80 &
   $M_{dark} / M_{lum}$ = 0.67 \\
 & $M_{(dark + lum)}$ = $1.3 \times 10^{10}$ \msol &
   $M_{(dark + lum)}$ = $1.2 \times 10^{10}$ \msol \\
 & & \\
At the last point: & $r_{outer}$ = 7.35 kpc & $r_{outer}$ = 6.84 kpc \\
 & $\rho_{halo}$ = 0.0015 \msol pc$^{-3}$ &
   $\rho_{halo}$ = 0.0015 \msol pc$^{-3}$ \\
 & $M_{dark} / M_{lum}$ = 0.97 &
   $M_{dark} / M_{lum}$ = 0.75 \\
 & $M_{(dark + lum)}$ = $1.5 \times 10^{10}$ \msol &
   $M_{(dark + lum)}$ = $1.3 \times 10^{10}$ \msol \\
\tableline\tableline
\end{tabular}
\tablenotetext{a}{\citet{car90}}
\tablenotetext{b}{this work}
\end{table}

\section{\ha Extent}

The first mention that NGC 7793 had an extended ionized component
was by \citet{mon71} from quite shallow photographic data. A deeper
and more detailed study was done more recently by \citet{fer96}.
They showed that the observed \ha luminosity of the DIG component of
NGC 7793 was large, accounting for $\sim$ 40\% of the total \ha
emission, similar to what is seen in M31 \citep{wal94} or the SMC
\citep{ken95}. \citet{fer96} argue that the integrated minimum Lyman
continuum (Lyc) power required to sustain the DIG is enormous and
can only easily be met by the ionizing output from massive star
formation; the mechanical energy from supernovae and stellar winds
falls short of that required by at least a factor of 2. Their
results support the hypothesis that the DIG is photoionized by Lyc
photons leaking out of discrete \hII regions.

\begin{figure}
%\centering
%\includegraphics[width=\textwidth]{haprof.eps}
\plotone{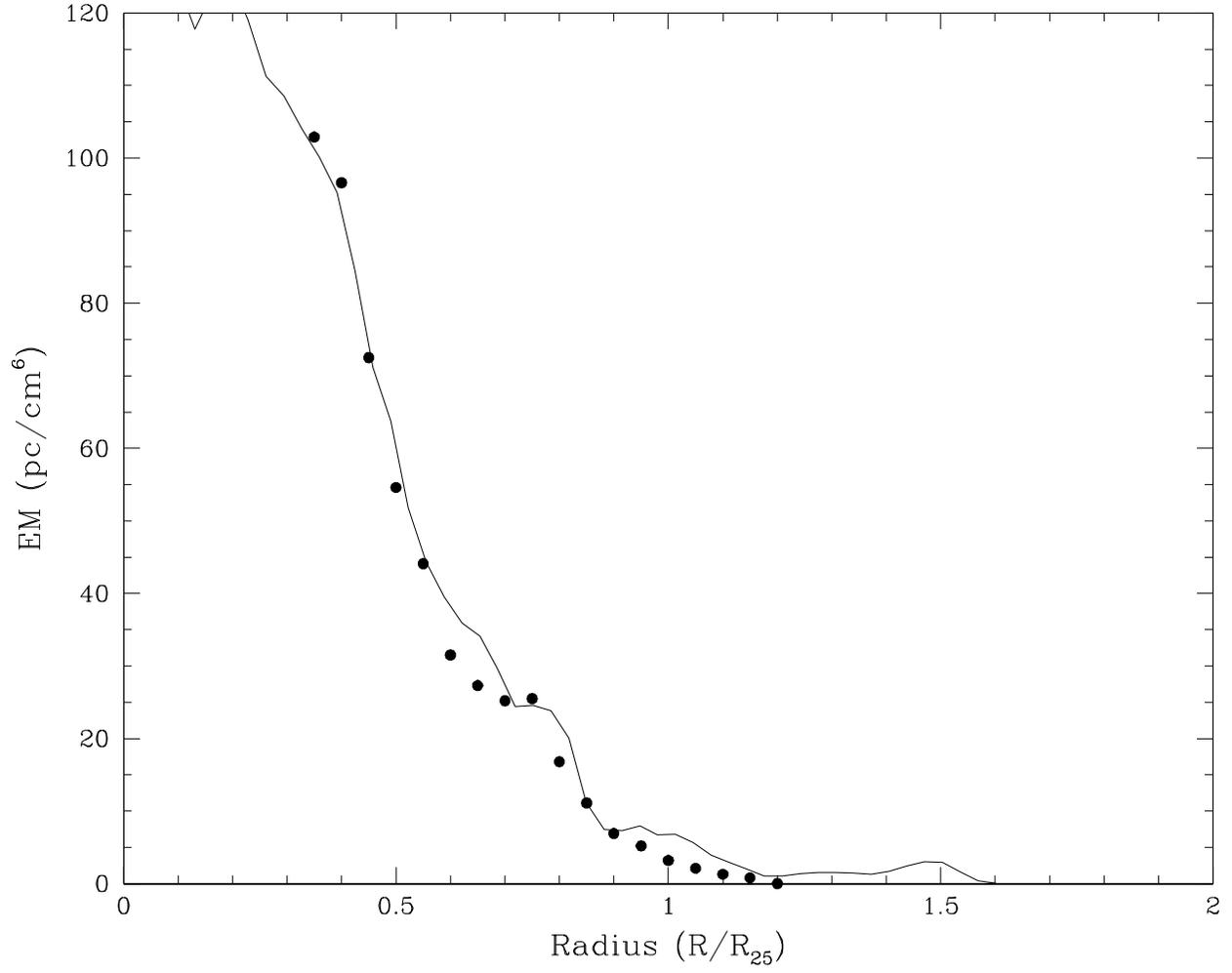}
\caption{Radial \ha profile of NGC 7793. Superposed on it is the radial profile
(filled circles) of \citet{fer96}.}
\label{plt_haprof}
\end{figure}

A comparison between our deep Fabry-Perot \ha map and the deep \ha
continuum subtracted image from Ferguson et al. shows a very good
general agreement. Nevertheless, we reach fainter diffuse \ha
emission in the outskirts of the galaxy located between \hII
regions. In addition, spurs are visible on our image that were not
detected by \citet{fer96}. Fig.~\ref{plt_haprof} shows the \ha
radial profile of NGC 7793, derived using the \textit{GIPSY} task
\textit{ELLINT} \citep{vog01}. Since our data were not
photometrically calibrated, our profile was adjusted in order to be
in rough agreement with the profile of \citet{fer96} in the outer
parts. This should be looked at as just an order of magnitude
representation since the \citet{fer96} data were obtained in imagery
with a 68\AA\ filter which included not only \ha but also the [\nII]
lines. While they corrected for the [\nII] contamination using
[\nII]/\ha = 0.22, \citet{bla97} found a ratio close to unity in the
outer parts of the other Sculptor Group galaxy NGC 253. So, even if
the intensity scale could be off by a factor of a few, this is
sufficient to show that our long Fabry-Perot observations (20 hours
total or 20 minutes per channel) on a 36 cm telescope nearly reach
the limit of the \hI disk \citep[$D_{HI} = 1.7 D_{25}$ ;][]{car90}.
To be able to reach so large radii is interesting since it was
estimated that if the truncation edge \citep{cor89, van93} of the
\hI disks is due to ionization of the hydrogen by the cosmic
background radiation, the expected flux \citep{bla94, bla97} should
be an \ha emission measure \E$_m$(\ha) $\sim 0.2-2.0$ pc cm$^{-6}$,
which is the level reached by our deep observations.

Contrary to what was found by \citet{bla97} (hereafter referred to
as BFQ1997) for NGC 253, the \ha disk of NGC 7793 does not extend
past the \hI emission even if it gets close to it. Indeed, BFQ1997
have detected ionized gas out to 1.4 $R_{25}$, while the \hI extends
only out to 1.2 $R_{25}$. We should nevertheless notice that these
authors did not give an \ha map of NGC 253 but only presented some
discrete \ha emission located at a single position. Using
Fabry-Perot techniques as well, they computed a radial velocity for
this emission region and since they used a fixed etalon, instead of
a scanning one as we did, they do not have full spatial coverage of
the galaxy. Moreover, they have only observed one side of the
galaxy.  Further observations of NGC 253 are needed to verify if the
outer \hII region detected by BFQ1997 is isolated or if it is a
fragment of a more extended region. In addition, it is necessary to
search for a counterpart of this region on the other side of the
galaxy to strengthen their conclusions.

\section{Discussion}

The outer regions of galaxies are expected to deliver important
clues on the way galaxies are assembled. It is into these regions
that material was accreted during violent interactions in the past
and where new material continues to fall today. Cosmological
simulations of disk galaxy formation predict a wealth of baryonic
structures and substructures up to ultra-faint surface brightness
levels, possibly due for instance to tidal debris from accretion and
disruption of an expected rich population of luminous satellites
\citep{bul05}.  The faintest features visible in M31 reach effective
V-band surface brightnesses greater than $\mu_V\sim30$
mag~arcsec$^{-2}$. The currently most powerful technique to probe
the very low surface brightness regions is resolved star counts, but
it is only applicable to our nearest large neighbors, such as M31
and M33. Both galaxies appear to show evidence for metal-poor,
pressure-supported stellar halos but only M31 shows evidence for
recent accretion \citep{iba05}. The combination of diffuse light
surface photometry with resolved star counts allowed \citet{irw05}
to reach an effective V-band surface brightness of $\mu_V\sim32$
mag~arcsec$^{-2}$ for M31. Globular cluster and planetary nebula
populations also offer additional clues to the formation and
evolution at large radii.

The presence of the neutral gas component, detected through its \hI
emission, at radii several times larger than the optical disk
(measured from its surface brightness at $\mu_B = 25$
mag~arcsec$^{-2}$) has now been well known for more than 30 years.
Many of the extended \hI rotation curves are rising or flat at their
outermost points, while others are slightly decreasing.
Nevertheless, none of them show a Keplerian decline, meaning that
the edge of the mass distribution has not yet been found
\citep{puc91}. The only suggestion that such an edge could have been
reached is for the dwarf galaxy DDO 154 \citep{car98}. Deep \hI
observations have shown that \hI disks are abruptly truncated at
column densities near $10^{18}-10^{19}$ atoms cm$^{-2}$, where the
rotation curves still do not have a Keplerian behavior (e.g. Van
Gorkom 1993).  If the metagalactic UV background is sufficiently
strong, beyond a certain galactic radius the cold gas may become
ionized \citep{boc77}

The total \ha emission from star forming galaxies is divided roughly
in half, between discrete \hII regions and the DIG. Pioneering work
by \citet{rey84} established that only the massive OB stars are able
to ionize the DIG.  On the other hand, all the observers agree on
the fact that the DIG fraction seems to be independent of galaxy
parameters such as star formation rate and Hubble type. This looks a
bit controversial. More recently, \citet{rey99} and \citet{col01}
suspected that OB stars cannot exclusively explain the ionization
state of the DIG. \citet{dop06} suggested that part of the DIG may
consist of extremely evolved filamentary \hII regions that may be
difficult to detect.

On the other hand, if ionizing radiation escapes from \hII regions
to ionize the DIG, it does escape from the galaxies and affects the
ionization state of the intergalactic medium (IGM) and the galactic
medium of external galaxies which may be partially at the origin of
this UV radiation. Indeed, a key question is the origin of the
ultra-violet photons responsible for ionizing the hydrogen in the
intergalactic medium (the Ly$\alpha$ forest). The most likely
sources of ionizing photons are young hot stars in galaxies and QSOs
\citep{bar01} but their exact origin remains one of the unsolved
questions of modern cosmology. Observations of high-redshift quasars
imply that the reionization of cosmic hydrogen was completed by
z$\sim$6 \citep[e.g.][and reference therein]{fan06}. Lack of
observational evidence and theoretical uncertainties makes it
difficult to estimate the contribution made by both sources and
leads to inconsistent results. To complicate the scheme, the
relative contribution of QSOs and star forming galaxies depends on
the lookback time \citep[and reference therein]{fan06}.

Recently, based on the first data release of the SINGG survey of
HI-selected galaxies \citep{meu06}, \citet{oey07} have undertaken a
quantification of the ionization in neutral galaxies and the diffuse
and warm ionized medium (NGC 7793 is unfortunately not included in
the first SINGG data release of 109 galaxies). They concluded that
the mean fraction of diffuse ionized gas in their sample is
$\sim$0.59, slightly higher but consistent with the value $\sim$0.40
found in previous samples \citep[e.g.][]{thi02, zur00, wan99, hoo96}
and also consistent with the value $\sim$0.50 found by \citet{fer96}
for NGC 7793.  The difference between \citet{oey07} and previous
studies may be due to selection effects favoring large, optically
bright and nearby galaxies with high star formation rates.  NGC 7793
is closer than any (except one) of the first SINGG data release.  We
have computed, for NGC 7793,  its \ha effective (half light) radius
$\sim1.97\pm0.05$ kpc ($\sim120$ arcsec).  Refering to equation (1)
from \citet{oey07}, and using the \ha total luminosity of
$4.1~10^{40}$ erg.s$^{-1}$ computed by \citet{fer96}, we found that
the logarithm of effective surface brightness
Log($\Sigma_{H{\alpha}}$) = 38.65 Log(erg~s$^{-1}$~kpc$^{-2}$.
Following \citet{oey07}, this brings NGC 7793 to be classified as a
normal galaxy with respect to its star formation rate.

\subsection{Possible Origins of the Ionization: Internal and/or External}

Several mechanisms for the ionization of the neutral hydrogen and
for heating the electron population belonging to the DIG may be
found in the literature. BFQ1997 review them in the specific context
of NGG 253.  The main source of ionization and heating may be (see
references in BFQ1997): the metagalactic ionizing background;
compact halo sources; compact disk sources (white dwarf populations
or horizontal-branch stars, young enough to produce significant UV
emission); ram pressure heating (NGC 253 and NGC 7793 lie both in
the Sculptor group having little intracluster medium);
turbulence-driven MHD-wave heating; mixing layers driven by bulk
flows; galactic fountain material driven by the inner starburst
through shocks; dilute photoionization; gas phase depletion (induced
by the formation of grains); low energy cosmic-ray electrons and
young stellar disk.

BFQ1997 ruled out most of these mechanisms, except the ionization by
the hot young stars of the disk. Even if those young stars are in
the disk and most of them further in, the warp of the \hI disk
(which is the case for both NGC 253 and NGC 7793) makes it possible
to expose it to their radiation. In particular, they ruled out the
ionization by the cosmic UV background which is suspected to be
$\sim$5 times below the ionizing flux necessary to explain the
emission seen in NGC 253. As the \ha disk does not reach the
outskirts of the \hI disk of NGC 7793, it is very unlikely that the
source of ionization of the DIG is the metagalactic UV background.
Indeed, if the metagalactic UV background was at the origin of the
ionization it should have principally affected the outer \hI
regions, which is not the case.  An alternative could be that the
\ha detection level is not yet low enough to be able to detect the
ionization by the cosmic UV background radiation, thus, this should
mean that the expected flux should have an \ha emission measure
\E$_m$(\ha) lower than what was measured for NGC 7793, namely $\sim
0.2-2.0$ pc cm$^{-6}$.

\subsection{A Truly Declining Rotation Curve}

The existence of falling RCs was first claimed in the context of the
effect of cluster environments on the member galaxies. Following the
work of \citet{dre80}, the increase of intergalactic gas and of
gravitational interactions expected toward the centers of clusters
should not only affect the morphological types of the systems but
also their kinematics and mass distribution. \citet{whi88}, from
long-slit emission-line observations, presented evidence for a
correlation in the sense that the inner-cluster spirals tend to have
falling RCs, while those of the outer cluster and the vast majority
of field spirals seem to have flat or even rising RCs. However,
\citet{amr93}, using full 3-D Fabry-Perot data did not confirm these
correlations and moreover found that many galaxies for which
\citet{whi88} claimed a falling RC, had in fact a flat or even
rising RC after a careful analysis of the Fabry-Perot 2D velocity
fields \citep{amr96}.

Declining RCs have been seen before in other systems \citep{cas91,
hon97, ryd98, sof01, noo07}. However, those galaxies are most of the
time earlier type systems (S0 - Sbc) with highly-concentrated
stellar light distribution and the maximum rotation velocity is
reached at small radii (few hundreds kpc and $< \alpha^{-1}$). This
is certainly not the case for NGC 7793 which has a very small bulge
and a relatively diffuse (exponential) light distribution and where
$V_{max}$ is reached at $\sim 4 \alpha^{-1}$. Keplerian declines
have also been observed in a few edge-on systems (e.g. NGC 891:
\citet{san79, oos07}, NGC 3079: \citet{vei95} \& NGC 4244:
\citet{oll96}), but the interpretation in the case of those galaxies
is much more difficult. Finally, a truncation of the halo at $\sim
10$ kpc was advocated to explain the decline of the last two points
of the RCs of NGC 5204 \citep{sic97} and NGC 253 \citep{bla97}, but
here the decrease of $V_{max}$ and the last velocity point is only
$\sim 10$\% of $V_{max}$ and $\sim 10$\% of $r_{max}$, while, for
NGC 7793, the decrease is $\sim 25$\% of $V_{max}$ and over about
half the radius range.

So, for late-type nearly bulgeless galaxies, the decreasing RC of
NGC 7793 is rather unique. As can be seen in Fig.~\ref{vrotfinal},
the decline of the RC seen with the \hI data is confirmed by the \ha
36cm data and the kinematical analysis of the 3 sets of data (\hI \&
\ha) give very similar results. Since a bulge component can not be
invoked to explain the falling gradient of the RC, NGC 7793 really
seems to have a genuine declining RC. The new data also confirm that
the decline is slower than Keplerian and that a dark component is
still necessary to explain the observed kinematics.

\section{Summary and Conclusions}

Deep \ha observations of NGC 7793
have been presented. The main results are:

\begin{itemize}
\item A total of 20 hours of observations was obtained on the
Marseille 36cm telescope which allowed to reach sensitivities of the
order of \E$_m$(\ha) $\sim 0.2-2.0$ pc cm$^{-6}$;
\item this allowed the detection of diffuse \ha emission out to the
edge of the \hI disk;
\item a warp of the disk plane, already seen in the \hI data
is also seen in the \hII disk;
\item a RC is derived out to 2.2\arcmin\ from the 3.6m data and
out to 7\arcmin\ from the 36cm data;
\item the two \ha RCs superposed exactly with the \hI RC;
\item the \ha results confirm the \hI findings that NGC 7793
has a truly declining rotation curve;
\item a model of the mass distribution gives a total mass for
NGC 7793 of $M_{(dark + lum)}$ = $1.3-1.5 \times 10^{10}$ \msol for
$M_{dark} / M_{lum}$ = 0.75-1.0 at the last velocity point.
\end{itemize}

This was our first experiment of doing such long integrations using
a 36cm telescope coupled with our \FM\ scanning Fabry-Perot system.
This yielded very interesting results for NGC 7793 by detecting
diffuse \ha emission all the way to the edge of the \hI disk and
confirming its peculiar kinematics. In the near future, we intend to
carry out the same kind of observations for all the Sculptor Group
galaxies, hoping to reach even fainter levels. This should be
possible with an even more sensitive camera soon available
\citep{dai04, dai06c}, based on an EMCCD, which should allow us to
increase the detective quantum efficiency by a factor $\sim$ 3-4.

\acknowledgments We would like to thank the staff of the ESO La
Silla Observatory for their support. We acknowledge support from the
Natural Sciences and Engineering Research Council of Canada and the
Fonds Qu\'eb\'ecois de la recherche sur la nature et les
technologies. The Digitized Sky Surveys (DSS images) were produced
at the Space Telescope Science Institute under U.S. Government grant
NAG W-2166. The images of these surveys are based on photographic
data obtained using the Oschin Schmidt Telescope on Palomar Mountain
and the UK Schmidt Telescope. The plates were processed into the
present compressed digital form with the permission of these
institutions. The IR images were obtained by the Spitzer Space
Telescope, which is operated by the Jet Propulsion Laboratory,
California Institute of Technology under a contract with NASA.


\clearpage

\begin{thebibliography}{}
\bibitem[Amram et al.(1993)]{amr93} Amram, P., Sullivan III, W. T.,
    Balkowski, C., Marcelin, M. \& Cayatte, V. 1993, \aap, 403, L59
\bibitem[Amram et al.(1996)]{amr96} Amram, P., Balkowski, C.,
    Boulesteix, J., Cayatte, V., Marcelin, M. \&
    Sullivan III, W. T. 1996, \aap, 310, 737
\bibitem[Barkana \& Loeb(2001)]{bar01} Barkana, R. \& Loeb, A. 2001,
    \physrep, 349, 125
\bibitem[Bland--Hawthorn et al.(1994)]{bla94} Bland--Hawthorn, J.,
    Taylor, K., Veilleux, S. \& Shopbell, P. L. 1994, \apj, 437, L95
\bibitem[Bland--Hawthorn et al.(1997)]{bla97} Bland--Hawthorn, J.,
    Freeman, K. C. \& Quinn, P. J. 1997, \apj, 490, 143 (BFQ1997)
\bibitem[Bochkarev \& Siuniaev(1977)]{boc77} Bochkarev, N. G. \&
    Siuniaev, R. A. 1977, \sovast, 21, 542
\bibitem[Boulesteix et al.(1984)]{bou84} Boulesteix, J., Georgelin,
    Y. P., Marcelin, M. \& Monnet, G. 1984, \procspie, 445, 37
\bibitem[Bullock \& Johnston(2005)]{bul05} Bullock, J. \& Johnstone, K.
    2005, \apj, 635, 931
\bibitem[Carignan(1985a)]{car85a} Carignan, C. 1985a, \apjs, 58, 107
\bibitem[Carignan(1985b)]{car85b} Carignan, C. 1985b, \apj, 299, 59
\bibitem[Carignan \& Freeman(1985)]{car85c} Carignan, C. \& Freeman, K. C.
    1985, \apj, 294, 494
\bibitem[Carignan \& Puche(1990)]{car90} Carignan, C. \& Puche, D.  1990,
    \aj, 100, 394
\bibitem[Carignan \& Purton(1998)]{car98} Carignan, C. \& Purton, C.
    1998, \apj, 156, 125
\bibitem[Casertano \& van Gorkom(1991)]{cas91} Casertano, S. \& van
    Gorkom, J. H. 1991, \aj, 101, 1231
\bibitem[Chemin et al.(2006)]{che06} Chemin, L., Balkowski, C., Cayatte, V.,
    Carignan, C., Amram, P., Garrido, O., Hernandez, O., Marcelin, M.,
    Adami, C., Boselli, A. \& Boulesteix, J. 2006, \mnras, 366, 812
\bibitem[Collins \& Rand(2001)]{col01} Collins, J. A. \& Rand, R. J.
    2001, \apj, 551, 57
\bibitem[Corbelli et al.(1989)]{cor89} Corbelli, E., Schneider, S. E.
    \& Salpeter, E. E. 1989, \aj, 97, 390
\bibitem[Daigle et al.(2004)]{dai04} Daigle, O., Gach, J.-L.,
    Guillaume, C., Balard, P. \& Boissin, O. 2004, \procspie, 5499, 219
\bibitem[Daigle et al.(2006a)]{dai06a} Daigle, O., Carignan, C., Amram, P.,
    Hernandez, O., Chemin, L., Balkowski, C. \& Kennicutt, R. 2006a,
    \mnras, 367, 469
\bibitem[Daigle et al.(2006b)]{dai06b} Daigle, O., Carignan, C.,
    Hernandez, O., Chemin, L. \& Amram, P. 2006b, \mnras, 368, 1016
\bibitem[Daigle et al.(2006c)]{dai06c} Daigle, O., Carignan, C. \&
    Blais--Ouellette, S. 2006c, in High Energy, Optical, and Infrared
    Detectors for Astronomy II. Eds David A. Dom and Anrew D. Holland,
    \procspie, 6276, 62761F
\bibitem[Davoust \& de Vaucouleurs(1980)]{dav80} Davoust, E. \& de
    Vaucouleurs, G. 1980, \apj, 242, 30
\bibitem[Dettmar \& Schulz(1992)]{det92} Dettmar, R. J. \& Schulz, H. 1992,
    \aap, 254, L25
\bibitem[de Vaucouleurs(1959)]{dev59} de Vaucouleurs, G. 1959, \apj, 130, 718
\bibitem[de Vaucouleurs et al.(1991)]{dev91} de Vaucouleurs, G.,
    de Vaucouleurs, A., Corwin, H. G. Jr., Buta, R. J.,
    Paturel, G. \& Fouqu\'e, P. 1991, Third Reference Catalogue
    of Bright Galaxies (New--York: Springer)
\bibitem[Dopita et al.(2006)]{dop06} Dopita, M., Fischera, J.,
    Sutherland, R. S., Kewley, L. J., Tuffs, R. J., Popescu, C. C.,
    van Breugel, W., Groves, B. A. \& Leitherer, C. 2006,
    ApJ, 647, 244
\bibitem[Dressler(1980)]{dre80} Dressler, A. 1980, \apj, 236, 351
\bibitem[Fan et al.(2006)]{fan06} Fan, X., Strauss, M. A., Becker, R. H.,
    White, R. L., Gunn, J. E., Knapp, G. R., Richards, G. T.,
    Schneider, D. P., Brinkmann, J. \& Fukugita, M. 2006, \aj, 132, 117
\bibitem[Ferguson et al.(1996)]{fer96} Ferguson, A. M. N., Wyse, R. F. G.,
    Gallagher III, J. S. \& Hunter, D.A. 1996, \aj, 111, 2265
\bibitem[Gach et al.(2002)]{gac02} Gach, J.--L., Hernandez, O.,
    Boulesteix, J., Amram, P., Boissin, O., CArignan, C.,
    Garrido, O., Marcelin, M., \"Ostlin, G., Plana, H.
    \& Rampazzo, R. 2002, \pasp, 114, 1043
\bibitem[Hernandez et al.(2003)]{her03} Hernandez, O., Gach, J.--L.,
    Carignan, C. \& Boulesteix, J. 2003, in Instrument Design and
    Performance for Optical/Infrared Ground-based Telescopes,
    Eds Iye, M. \& Moorwood, A. F. M., Proceedings of the SPIE, 4841, 1472
\bibitem[Hernandez et al.(2005)]{her05} Hernandez, O., Carignan, C.,
    Amram, P., Chemin, L. \& Daigle, O. 2005, \mnras, 360, 1201
\bibitem[Honma \& Sofue(1997)]{hon97} Honma, M. \& Sofue, Y. 1997,
    \pasj, 49, 539
\bibitem[Hoopes et al.(1996)]{hoo96} Hoopes, C. G., Walterbos, R. A. M.
    \& Greenawalt, B. E. 1996, \aj, 112, 1429
\bibitem[Ibata et al. (2005)]{iba05} Ibata, R., Chapman, S., Ferguson,
    A. M. N., Lewis, G., Irwin, M. \& Tanvir, N. 2005, \apj, 634, 287
\bibitem[Irwin et al.(2005)]{irw05} Irwin, M. J., Ferguson, A. M. N.,
    Ibata, R. A., Lewis, G. F. \& Tanvir, N. R. 2005, \apj, 628, L105
\bibitem[Karachentsev et al.(2003)]{kar03} Karachentsev, I. D. 2003,
    \aap, 404, 93
\bibitem[Karachentsev(2005)]{kar05} Karachentsev, I. D. 2005, \apj, 129, 178
\bibitem[Kennicutt et al.(1995)]{ken95} Kennicutt, R. C., Bresolin, F.,
    Bomans, D. J., Bothun, G. D., \& Thompson, I. B. 1995, \aj, 109, 594
\bibitem[Meurer et al.(2006)]{meu06} Meurer, G. R. et al. 2006, \apjs, 165, 307
\bibitem[Monnet(1971)]{mon71} Monnet, G. 1971, \aap, 12, 379
\bibitem[Noordermeer(2007)]{noo07} Noodermeer, E., van der Hulst, J. M.,
    Sancisi, R., Swaters, R. S. \& van Albada, T. S. 2007, \mnras, 376, 1513
\bibitem[Oey et al.(2007)]{oey07} Oey, M. S. et al. 2007, \apj, 661, 801
\bibitem[Oosterloo et al.(2007)]{oos07} Oosterloo, T., Fraternali, F.
    \& Sancisi, R. 2007, \aj, 134, 1019
\bibitem[Olling(1996)]{oll96} Olling, R. P. 1996, \aj, 112, 457
\bibitem[Puche \& Carignan(1988)]{puc88} Puche, D. \& Carignan, C. 1988,
    \aj, 95, 1025
\bibitem[Puche \& Carignan(1991)]{puc91} Puche, D. \& Carignan, C. 1991,
    \apj, 378, 487
\bibitem[Rand et al.(1990)]{ran90} Rand, R. J., Kulkarni, S. R. \&
    Hester, J. J. 1990, \apj, 352, 1
\bibitem[Reynolds(1984)]{rey84} Reynolds, R. J. 1984, \apj, 378, 487
\bibitem[Reynolds et al.(1999)]{rey99} Reynolds, R. J., Haffner, L. M. \&
    Tufte, S. L. 1999, \apj, 525, L21
\bibitem[Ryder et al.(1998)]{ryd98} Ryder, S. D., Zasov, A. V.,
    McIntyre, V. J., Walsh, W. \& Sil'chenko, O. K. 1998, \mnras, 293, 411
\bibitem[Sancisi \& Allen(1979)]{san79} Sancisi, R. \& Allen, R. J. 1979,
    \aap, 74, 73
\bibitem[Sicotte \& Carignan(1997)]{sic97} Sicotte, V. \& Carignan, C.
    1997, \aj, 113, 609
\bibitem[Sofue \& Rubin(2001)]{sof01} Sofue, Y. \& Rubin, V. 2001,
    \araa, 39, 137
\bibitem[Thilker et al.(2002)]{thi02} Thilker, D. A., Walterbos, R. A. M.,
    Braun, R. \& Hoopes, C. G. 2002, \aj, 124, 3118
\bibitem[van Gorkom(1993)]{van93} van Gorkom, J. H. 1993, in The
    Environment and Evolution of Galaxies, ed. J.M. Shull \& H.A.
    Thronson (Dorfrecht: Kluwer), p. 345
\bibitem[Veilleux et al.(1995)]{vei95} Veilleux, S., Cecil, G. \&
    Bland--Hawthorn, J. 1995, \apj, 445, 152
\bibitem[Veilleux et al.(2003)]{vei03} Veilleux, S., Shopbell, P. L.,
    Bland--Hawthorn, J. \& Cecil, G. 2003, \apj, 126, 2185
\bibitem[Vogelaar \& Terlouw(2001)]{vog01} Vogelaar, M. G. R. \&
    Terlouw, J. P. 2001, in Astronomical Data Analysis Software and
    Systems X, eds F. R. Harnden Jr., F. A. Primiri \& H. E. Payne,
    ASP Conf. Series, 238, 358
\bibitem[Walterbos \& Braun(1994)]{wal94} Walterbos, R. A. M. \& Braun, R.
    1994, \apj, 431, 156
\bibitem[Wang et al.(1999)]{wan99} Wang, J., Heckman, T. M. \& Lehnert, M. D.
    1999, \apj, 515, 97
\bibitem[Whitmore et al.(1988)]{whi88} Whitmore, B. C., Forbes, D. A.
    \& Rubin, V. C. 1988, \apj, 333, 542
\bibitem[Zurita et al.(2000)]{zur00} Zurita, A., Rozas, M. \& Beckman, J. E.
    2000, \aap, 363, 9
\end{thebibliography}
\end{document}